\documentclass{article}\usepackage{jkas2}
\runningauthor{G.Brunetti}
\runningtitle{
Particle acceleration and non--thermal emission from galaxy clusters
}
\beginpage{1}
\endpage{5}

\begin{document}

\font\twelvei = cmmi10 scaled\magstep1 
       \font\teni = cmmi10 \font\seveni = cmmi7
\font\mbf = cmmib10 scaled\magstep1
       \font\mbfs = cmmib10 \font\mbfss = cmmib10 scaled 833
\font\msybf = cmbsy10 scaled\magstep1
       \font\msybfs = cmbsy10 \font\msybfss = cmbsy10 scaled 833
\textfont1 = \twelvei
       \scriptfont1 = \twelvei \scriptscriptfont1 = \teni
       \def\mit{\fam1 }
\textfont9 = \mbf
       \scriptfont9 = \mbfs \scriptscriptfont9 = \mbfss
       \def\bmit{\fam9 }
\textfont10 = \msybf
       \scriptfont10 = \msybfs \scriptscriptfont10 = \msybfss
       \def\bmsy{\fam10 }

\def\etal{{\it et al.~}}
\def\eg{{\it e.g.,~}}
\def\ie{{\it i.e.,~}}
\def\lsim{\raise0.3ex\hbox{$<$}\kern-0.75em{\lower0.65ex\hbox{$\sim$}}}
\def\gsim{\raise0.3ex\hbox{$>$}\kern-0.75em{\lower0.65ex\hbox{$\sim$}}}

\title{Particle acceleration and non--thermal emission from galaxy
clusters}

\author{Gianfranco Brunetti $^{1}$}
\address{$^{1}$ Istituto di Radioastronomia del CNR/INAF,
via P.Gobetti 101, Bologna, I--40129, Italy}

%



\abstract{
The existence and extent of non--thermal phenomena in galaxy
clusters is now well established.
A key question in our understanding of these phenomena is 
the origin of the relativistic electrons which 
may be constrained by the modelling of the fine radio 
properties of radio halos and of their statistics.
In this paper we argue that present data favour a scenario in which
the emitting electrons in the intracluster medium 
(ICM) are reaccelerated {\it in situ} on their way out.
An overview of turbulent--particle acceleration models is given focussing
on recent time--dependent calculations which include a full coupling between 
particles and MHD waves.}


\maketitle

\section {INTRODUCTION}

There is now firm evidence that the ICM is a mixture of hot gas, 
magnetic fields and relativistic particles. 
While the hot gas results in thermal
bremsstrahlung X-ray emission, relativistic electrons
and positrons generate non-thermal radio (synchrotron) and 
possibly hard X-ray radiation (inverse Compton, IC).
In principle, relativistic hadrons can store a relevant fraction 
of the energy budget of the ICM since they are 
essentially no--loss
particles and remain confined in the cluster volume 
(V\"{o}lk et al. 1996;
Berezinsky, Blasi \& Ptuskin 1997).
However, the gamma radiation that would allow us to
constrain the energetics of these particles
has not been detected as yet (Reimer et al., 2003).
Our understanding of the non-thermal activity in 
galaxy clusters is thus based on the synchrotron
and IC emission from the relativistic
leptons.
The most spectacular example 
is given by the giant radio halos (e.g.
Feretti 2003).
The difficulty in explaining these sources arises from the
combination of their $\sim$Mpc size, and the relatively short radiative
lifetime of the radio emitting electrons (about $10^8$yrs):
the diffusion time necessary to these electrons to cover such 
distances is much larger than their radiative lifetime.
This is still an open problem and
two main possibilities have been proposed so far.
Relativistic electrons can be continuously reaccelerated 
{\it in situ} on their way out (e.g., Jaffe 1977), or 
these electrons can be continuously injected in the 
ICM by inelastic proton-proton collisions through
production and decay of charged pions (secondary models;
e.g. Dennison 1980).

After a discussion on the observations which 
seem to favour the first hypothesis, we focus
on particle acceleration models; H$_o$=50 km/s/Mpc
is assumed.

\section {SECONDARY MODELS \& OBSERVATIONS}

A relatively straightforward possibility to have 
synchrotron
emission on Mpc scale in galaxy clusters is provided by a continuous
injection of fresh relativistic secondary 
electrons generated by hadronic collisions
(Dennison 1980; Blasi \& Colafrancesco 1999).
It has been shown that secondary models may be able to account for
the basic properties of the observed radio halos provided that the
strength of the magnetic field averaged over the emitting volume
is of the order of few $\mu$G or greater
(e.g., Dolag \& Ensslin 2000; Miniati et al.~2001).
On the other hand, it has been shown that 
the detailed properties are difficult to be explained through
these models 
(Brunetti 2003; Feretti et al.~2004; Kuo et al.~2004; Reimer
et al.~2004) suggesting that additional mechanisms of
particle acceleration should be active in the ICM.

In this Section we discuss in detail
some of these fine properties which thus become the tool
to discriminate among different explenations for the
origin of radio halos.
We show that current observations, 
although may be still affected by biases and incompleteness which 
deserve future studies, provide evidence in favour of 
diffuse electron acceleration in galaxy clusters.

\subsection {Basic Formalism for secondary electrons}

In order to derive general expectations to be compared 
with observations, in this Section we derive the basic
properties of the synchrotron emission from secondary
models.

Relativistic hadrons might be efficiently injected
in the ICM by cosmological shocks (e.g., Miniati
et al.~2001), Galactic Winds (e.g., V\"olk \& Atoyan
1999), and AGNs.
These particles do not suffer efficient energy losses
so that their spectrum can be described by a 
power law in momentum which extends up to very high
energies (e.g., Blasi 2001; see also Drury, and 
Blasi, these proceedings) :

\begin{equation}
N_{\rm p}(p)
= K_{\rm p} p^{-s}
\label{np}
\end{equation}

\noindent
and the energy density in the ICM is given by :

\begin{equation}
{\cal E}_{\rm p}
=
m_{\rm p} c^2 \int_{p_{\rm min}}
dp_{\rm p}
N_{\rm p}(p) \left[
\left(
1+ ( {{p_{\rm p}}\over{m_{\rm p}c}} )^2 
\right)^{ {1\over 2} }
-1
\right]
\label{ep}
\end{equation}

\noindent
where $p_{\rm min}$ is the minimum momentum of the protons which
are accelerated.
It is believed that shocks may play a leading role in the acceleration
of cosmic ray protons in the ICM.
In general, the injection of these particles at shocks is computed according
to the thermal leakage model (Kang \& Jones 1995) so that
the momentum threshold for the injected protons
is $p_{\rm min}=c_1 2(m_{\rm p} k_{\rm B} T_{\rm psh})^{1/2}$
where $k_{\rm B}$ and $T_{\rm psh}$ are the Boltzmann's constant and the post
shock temperature, respectively.
The number of thermal protons passing through shocks which are accelerated
at supra--thermal and relativistic energies is controlled by the value
of the parameter $c_1$.
Numerical simulations aimed at the modelling of non--thermal phenomena
in galaxy clusters adopt $c_1=2.6$ which is claimed to be consistent
with observational and theoretical studies
(e.g., Miniati et al.~2001 and ref. therein).

The inelastic collisions of relativistic protons with the thermal 
protons may generate a population of
secondary electrons due to the decay of charged pions.
The process of pion production in {\it pp} scattering is a threshold
reaction that requires protons with kinetic energy larger than $\sim$ 300
MeV.
The spectral distribution of the injection rate of secondary
electrons is given by (e.g., Mannheim \& Schlickeiser 1994) :

\begin{equation}
q_{\rm e}(p)
= C(s) n_{\rm th} K_{\rm p}
p^{- {2 \over 3} (2s-1) }
\label{qe}
\end{equation}

\noindent
and the spectrum of secondary electrons in the ICM can be estimated
under stationary conditions (e.g., Dolag \& Ensslin 2000):

\begin{equation}
N_{\rm e}(p)=
{1\over{ | \dot{p} |}}
\int_p^{\infty}
q_{\rm e}(p) dp
\label{ne}
\end{equation}

where

\begin{equation}
\dot{p}
= C_{\rm rad}
p^2
\left(
B^2 +
B_{\rm IC}^2(z)
\right)
\label{loss}
\end{equation}

\noindent
is the rate of energy losses (e.g.,
Sarazin 1999; Brunetti 2003) of the
injected electrons (for $\gamma \geq 10^3$);
$B_{\rm IC} \simeq 3.2 \mu$G$(1+z)^2$
and $C_{\rm rad}$ is a constant.
Combining Eqs.(\ref{qe},\ref{ne},\ref{loss}) one
has the spectrum of the secondary electrons :

\begin{equation}
N_{\rm e}(p)
=
{{
C_{\rm rad} C(s) n_{\rm th} K_{\rm p} 
}\over{
{4\over 3}s - {5\over 3}
}}
{{
p^{-({4 \over 3}s + {1\over 3} )}
}\over{
B^2 + B_{\rm IC}^2(z)
}}
= m_{\rm e} c K_{\rm e} \gamma^{-\delta}
\label{nefin}
\end{equation}

\noindent
The value of the maximum energy of the injected electrons is driven by 
that of the maximum energy of the accelerated protons which depends
on many unknown quantities.
However this
energy should be very high and thus the maximum 
energy of the injected secondary electrons is expected 
at energies well beyond those of the electrons emitting
synchrotron radiation in the radio band (e.g., Blasi 2001).

\subsection {Synchrotron spectra of Radio Halos}

For a power law energy distribution which extends
up to very high energies (Eq.\ref{nefin}),
the synchrotron spectrum is given by :

\begin{equation}
j_{\rm syn}
=
C_{\rm syn}(\delta)
K_{\rm e}
B^{\alpha_{\rm syn}+1}
\nu^{-\alpha_{\rm syn}}
\label{jsyn}
\end{equation}

\noindent
where $\alpha_{\rm syn}=(\delta-1)/2$, and $C_{\rm syn}(\delta)$ is 
given in classical books (e.g., Rybicki \& Lightman, 1979).
Assuming a strength of the magnetic field in
the ICM, from Eqs.(\ref{ep},\ref{nefin},\ref{jsyn}) 
it is possible to estimate the energy density 
of cosmic ray protons which is necessary to produce 
a fixed synchrotron emissivity (at $\nu$) due to the
secondary electrons :

\begin{equation}
{\cal E}_{\rm p}=
{{ C_{\rm E} C_{\rm rad} C(s) }\over
{C_{\rm syn}(\delta) }}
I(s,p_{\rm min})
{{ B^2 + B_{\rm IC}^2(z) }\over
{ B^{1+\alpha_{\rm syn}}}}
j_{\rm syn}(\nu) \nu^{\alpha_{\rm syn}}
\label{epsyn}
\end{equation}

\noindent
where $C_{\rm E}$ is a constant and $I()$ is the integral
in Eq.~(\ref{ep}).
In Fig.~1 we report the energy density of cosmic ray protons
which is required for a fixed $j_{\rm syn}$ (at 1.4 GHz) as 
a function of $\alpha_{\rm syn}$.
The hadrons' energy density is normalized to the case of a
Coma--like spectral index, $\alpha_{\rm syn}\simeq 1.14$.
With increasing $\alpha_{\rm syn}$ the fraction of secondary electrons
with $\gamma \sim 10^4$ (which emit at $\nu=$1.4 GHz) decreases and 
thus the total number (and energy) of these electrons and of the
relativistic protons should increase 
to match the fixed synchrotron flux.
This is dramatic for $\alpha_{\rm syn}> 1.5$ since in this
case the spectrum of protons is steep and most of the energy
is associated to the non--relativistic part of the particle spectrum.
From an energetical point of view, Fig.~1 basically 
indicates the existence of a forbidden
region for secondary models 
for values of the synchrotron spectral index $\alpha_{\rm syn} > 1.5$.
For example, for a fixed emissivity 
at 1.4 GHz secondary models with $\alpha_{\rm syn}=1.5$,
$1.7$ and $1.8$ should require an energy budget of relativistic
hadrons $\sim 10$, $50$ and $100$ times larger than 
that of a $\alpha_{\rm syn}= 1.14$ model, respectively.
In Fig.~1 we report the observed spectral indices of a sample
of well studied radio halos (Feretti et al.~, these
proceedings): half of the sample lies in the forbidden region.
Since most of these halos are more powerful and extended
than Coma C (for which $\geq$10\% of the
thermal energy should be required in the form of protons,
see above), the explenations of these radio 
halos may represent a serious challenge for secondary models.

A second point is given by the integrated synchrotron spectra.
In a number of well studied radio halos the spectrum steepens
at $\geq$ 1 GHz frequencies (e.g., Coma C: Thierbach et al.~2003;
Abell 754: Fusco--Femiano et al.~2003;
Abell 1914: Komissarov \& Goubanov 1994; Abell 2319: Feretti et al.~1997).
Ensslin (2002) claimed that the steepening observed in the 
Coma halo, where the 2.7 and 5 GHz points fall a factor of $\sim 1.8$ 
and $\sim 3.3$ below the extrapolation from the lower frequency
data, may be significantly mitigated if the SZ decrement
by the thermal electrons of the cluster is taken into account 
(a cluster radius R=5$h^{-1}_{50}$Mpc was assumed). 
The radio halo however covers only a small fraction of the
cluster area (e.g., Thierbach et al.~2003).
In the Rayleigh--Jeans part of the CMB spectrum,
the SZ distorsion of the CMB integrated over the
area of the radio halo is a negative flux; we obtain 
(Brunetti et al.~in prep.):

\begin{equation}
F_{\rm sz}^{-}(\nu)
\simeq
{{ 16 \pi}\over{D^2}}
{{ k_{\rm B}^2 T_{\rm cmb} T_{\rm th} \sigma_{T}}\over
{ m_{\rm e} c^2}}
( {{ \nu}\over{c}} )^2
\int_0^{R_H}
dx \, x
\int_{x}^{R_V}
{{ dr \,r n_{\rm th}(r) }\over
{ \sqrt{r^2 - x^2} }}
\label{sz}
\end{equation}

\noindent
where $D$ is the distance of the cluster, and $R_H$ and $R_V$ are
the radius of the halo and the virial radius of the cluster,
respectively.
This gives a correction of only $\sim 20 \%$ of the flux measured at 
5 GHz (and negligible at 2.7 GHz) when applied to the case
of Coma C (this correction is 
within the errorbars reported in Thierbach et al.,
2003). A similar result is obtained by adopting the value of
the Compton parameter measured for the Coma cluster
(Reimer et al.~2004, Figs.1-2).
Thus provided that the radio data are correct, 
the observed high frequency spectral steepenings 
are most likely related to the presence of a cut--off at 
$\gamma \sim 10^4$ in the spectrum of the emitting electrons.
This represents a serious challenge
for secondary models.

A thirdt point is given by the spatial distribution of
the synchrotron spectral indices of radio halos.
It is well known that the 0.3--1.4 GHz spectral index map
of the Coma radio halo indicates a progressive steepening of the
spectrum from the center to the periphery 
(with $\alpha_{\rm syn} \sim 0.8$ to $\sim 2$;
Giovannini et al.~1993).
More recently Feretti et al.~(2004) have found a similar trend
in the case of the giant radio halos in Abell 665 and 2163
along undisturbed cluster regions and complex spectral patches in
coincidence with dynamically disturbed regions
(see also Feretti et al.~these proceedings).
These observations suggest that the shape of the spectrum 
of the emitting electrons is relatively complex even on
100--200 kpc scales.
On the other hand, the continuous injection
of secondary electrons in the ICM is expected to produce 
power law spectra rather independent from the spatial
location in the cluster (at least on $<<$Mpc scales) and thus 
these observations represent a challenge for the hypothesis 
of a secondary origin of that the emitting electrons.

\subsection {Synchrotron Brightness Profiles of Radio Halos}

Giant radio halos are very extended sources (up to 2--2.5 Mpc).
In some cases the radial profiles of the
synchrotron emission are broader
than those of the X--rays emitted from the hot gas 
(e.g., Govoni et al.~2001).
This basically suggests that the decrease of the synchrotron emissivity
at a given distance is smaller than that of the bremsstrahlung
emissivity, $j_{\rm th} \propto n_{\rm th}^2$, and thus
that the relativistic electrons are much broadly distributed than
the thermal particles (e.g., Brunetti 2003).
In case these electrons are secondaries, 
this should imply a very broad distribution
of the cosmic ray protons and may cause an energetic problem.
The energetic problem may be alleviated 
by assuming stronger magnetic fields combined with 
a more gentle radial decay of the field strength, 
or by searching for minimum energy
configurations of the particles and field distributions
(Pfrommer \& Ensslin 2004).

To test the hypothesis of a secondary origin of the emitting 
electrons with the observed radial profiles 
it is necessary to focus on the case of the most extended
and powerful radio halos.
The radial dependence of the projected synchrotron brightness from
secondary electrons is given by :

\begin{eqnarray}
b_{\rm syn}(\nu,y)=
{3\over 2}
\nu^{-\alpha_{\rm syn}}
{{
C_{\rm syn}(\delta)
C_{\rm rad} C(s)}\over{
{4\over 3} - {s\over 3} }}
{{
k_{\rm B} T_{\rm th} K_{\rm p}(0)}\over{
{\cal E}_{\rm p}(0) }}
\nonumber\\
\times
\int_y
{{ dR \, R\, X_{\rm th}^{\rm p}(R) }\over{
\sqrt{R^2 - y^2} }}
n_{\rm th}^2(R)
{{
B^{1+\alpha_{\rm syn}}(R) }\over{
 B^2(R)+B_{\rm IC}^2(z) }}
\label{bright}
\end{eqnarray}

\begin {figure}[t]
\vskip 0cm
\centerline{\epsfysize=8cm\epsfbox{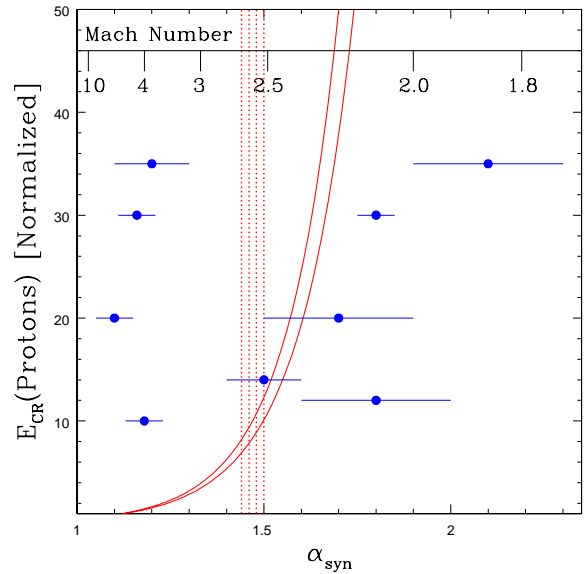}}
\vskip -0.3cm
\caption{Energy density (normalized) of cosmic ray protons necessary
to produce a fixed synchrotron flux at 1.4 GHz as a function 
of $\alpha_{\rm syn}$. The two curves are calculated for $B$= 1.0 (upper)
and 1.5 (lower) $\mu$G; $c_1=2.6$ is assumed. 
Data are taken from Feretti et al.~(these proc.).
The upper bar gives the Mach number of the shocks (if protons are
accelerated at shocks) necessary to give $\alpha_{\rm syn}$.}
\vskip -0.2cm
\end{figure}

\begin {figure}[t]
\vskip 0cm
\centerline{\epsfysize=8cm\epsfbox{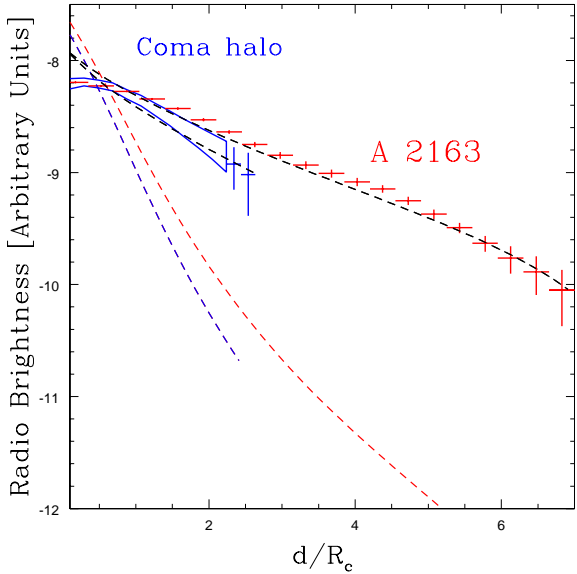}}
\vskip -0.3cm
\caption{Radial SYN profiles for Coma C (blue data) 
and for A 2163 (red data) as a function of the distance
from the center in units of the cluster core radius.
Secondary models are calculated to match the observed
radio luminosities. 
All the models adopt a central field $B_o=3 \mu$G.
Results are obtained assuming $B(r) \propto n_{\rm th}^{2/3}$
and $X_{\rm th}^{\rm p}$=const (red and blue dashed curves), and 
$B(r) \propto n_{\rm th}^{1/2}$ and $X_{\rm th}^{\rm p}$ increasing with 
distance (black curves).
The energy of relativistic protons is $\sim$4.5\% and
$\sim$17\% of the thermal pool 
in the case of the Coma cluster (blue and black curves,
respectively) and $\sim$60\% and $\sim$700\% for
A 2263 (red and black curves, respectively).
}
\vskip -0.2cm
\end{figure}

\noindent
where the parameter
$X_{\rm th}^{\rm p}(R)={\cal E}_{\rm p}(R)/{\cal E}_{\rm th}(R)$
gives the radial dependence of
the ratio between the energy densities
of the relativistic protons and of the
thermal particles.
A toy model which assumes a constant value of 
$X_{\rm th}^{\rm p}$ and a strength of the magnetic field,
$B \propto n_{\rm th}^{2/3}(R)$
(magnetic--flux freezing), gives brightness profiles 
which are much steeper than the observed ones (Fig.~2).
In order to reproduce the profiles
the parameter $X_{\rm th}^{\rm p}$ should increase with radius.
The result (with $X_{\rm th}^{\rm p}$ increasing with radius
and adopting a more gentle $B(R) \propto n^{1/2}_{\rm th}$), 
is reported in Fig.~2.
However, the large number of relativistic protons necessary 
to match the brightness in the external 
regions of the clusters inflates the energy budget associated 
to these particles: for example $\sim 7$ times the energy 
of the thermal pool is required 
in the case of Abell 2163. This is clearly unacceptable,
in addition, the energy density of the relativistic 
protons should increase by a factor $\sim 20$ going from
the core to the external regions of the cluster.
Obviously, the energetics of the relativistic protons may be reduced
by assuming a flatter profile of the magnetic field strength.
However sub--equipartition conditions (relativistic/thermal) 
in Abell 2163 and similar radio halos can be
reached only by adopting an almost flat
profile of the field strength on Mpc scales (Brunetti et al., 
in prep.), which however contrasts with the
current scenario in which the magnetic fields are strongly amplified 
in the ICM during cluster formation
(e.g., Dolag et al.~2004 and ref. therein).
Thus the explenation of the $>$ Mpc extension
of powerful radio halos (Abell 2163 class) is difficult 
if the emitting electrons are simply injected by
hadronic collisions.

\subsection {Connection with cluster mergers and occurrence of
Radio Halos}

Interestingly, there is a correlation between the non--thermal
diffuse radio emission and the presence of merger activity in
the host clusters of galaxies (Buote, 2001; Schuecker et al.~2001):
this suggests a link between the process of formation of
galaxy clusters and the origin of the non-thermal activity.
If this evidence is confirmed by future measurements
the secondary electron models have an additional
problem. In this case the radio emission would be dominated
at any time by the electrons continuously produced by the 
cosmic ray protons accumulated during the cluster history, 
rather than by the more recent merger events. 
No significant correlation should therefore be expected unless 
the decay time--scale of the magnetic fields amplified during a 
cluster merger on Mpc scale is much shorter than a Hubble time.
This however seems to be in contrast with the current scenario of 
a gradual amplification of magnetic fields in the ICM during
cluster formation (e.g., Dolag et al.~2004).

A very important point is the statistics of radio halos.
These sources are very rare if extracetd by a flux limited 
sample of galaxy clusters (Giovannini, Tordi, Feretti 1999).
However, the detection rate of these diffuse radio sources 
shows an abrupt increase with the X--ray luminosity of the host clusters:
about 30-35\% of the galaxy clusters with
X-ray luminosity larger than 10$^{45}$ erg s$^{-1}$
show diffuse radio emission
(Feretti, 2003).
Kuo et al.~(2004) have used this point as a tool to investigate
the origin of the emitting electrons and concluded that, given
the observed statistics, the typical life--time of radio halos
should be of the order of 1 Gyr, in contrast with a secondary 
origin of the emitting electrons which would produce very long
living radio halos (see also Kuo et al., these proceedings).
On the other hand, very recently 
it has been shown that an abrupt increase of the occurrence
of radio halos with the mass of the parent clusters can be
understood in the framework of particle acceleration models
(Cassano et al., these proceedings).

The importance of these points deserve 
additional observations to test the radio halos--
cluster merger connection and to better describe their statistics.

\section {PARTICLE ACCELERATION MODELS}

Given the difficulties which have secondary models 
in explaining the fine radio properties and the statistics 
of radio halos, one possibility is to admit the presence of particle
acceleration processes active in the ICM.

Although the physics of particle acceleration models is
difficult to be tested, it should be stressed that these models
predict very clear and general properties of radio halos 
which are almost independent from the details of the
adopted physics:

\begin{itemize}
\item[{\it i)}] in these models, the accelerated electrons
have a maximum energy at $\gamma < 10^5$ which produces a 
high frequency cut--off in the resulting
synchrotron spectrum (e.g., Schlickeiser et al.~1987;
Brunetti et al.~2001; Petrosian 2001).
This may naturally 
produce the observed steepenings of the synchrotron
spectrum and the complex spatial distributions 
of the synchrotron spectral index between two frequencies 
(Feretti et al.~2004);

\item[{\it ii)}]
a relatively thigh
connection of radio halos with cluster mergers 
is a very natural expectation of these models
(e.g., Tribble 1993: Roettiger et al.~1997).
In the framework of these models it can also be
shown that an abrupt increase of the possibility
to host giant radio halos with increasing mass
of the parent cluster is expected
(Cassano et al., these proceedings).
\end{itemize}

Obviously massive observations are necessary to 
confirm the spectral and statistical properties 
of radio halos, and thus to test the hypothesis of particle 
acceleration in the ICM.
In addition expectations from particle
acceleration models in the still unexplored observational
bands (e.g., $< 300$MHz and gamma rays) are required to 
efficiently test these models.

\subsection {CLUSTER MERGERS: TURBULENCE AND PARTICLE ACCELERATION}

It is well known that mergers and accretion of matter 
at the virial radius are likely to 
form shocks and to inject turbulent motions in the ICM
(e.g., Roettiger et al.~1997; Ricker \& Sarazin 2001).

There is still some debate on the typical Mach number 
of the shocks developed during cluster mergers 
(e.g., Miniati et al.~2001; Gabici \& Blasi 2003; Ryu et al.~2003) 
and on the resulting efficiency of particle acceleration in the
cluster volume (e.g., Ryu et al.~these meeting).
However, at least in the framework of {\it in situ}
particle acceleration
models, this is not relevant since the radiative
life--time of the emitting electrons is much shorter 
than the crossing time of sub--halos through the main clusters, 
and thus shock--acceleration cannot explain the scale and morphology 
of radio halos (e.g., Brunetti 2003).

Cluster mergers induce large--scale bulk 
flows with velocities $\sim 1000$ km s$^{-1}$ or larger.
The Kelvin--Helmotz instabilities generated during the passage
of the sub--clusters inject eddies which redistribute the energy 
of the mergers through the cluster volume 
and decay into turbulent velocity fields.
The numerical simulations find that an energy
budget of 10-30\% of the thermal energy of the ICM
can be channelled in the form of cluster turbulence (e.g.,
Sunyaev et al.~2003); also very
recent semi--analytical calculations
seem to be in rough agreement with these estimates 
(Cassano \& Brunetti 2004; Cassano \& Brunetti, these proceedings).
Spatially--resolved {\it pseudo--pressure} maps of the Coma cluster 
obtained from a mosaic of XMM--Newton observations have
revealed the signature of mildly supersonic turbulence 
(Schuecker et al. 2004).
The calorimeters onboard of 
ASTRO-E2 should be able
to detect the turbulent broadening of the lines of heavy ions
in excess to the thermal broadening (Inogamov \& Sunyaev 2003),
thus directly probing cluster turbulence.
There are a number of mechanisms to channel the energy flux of the
turbulence in the acceleration of fast particles.
Commonly used are acceleration 
via Magneto-Sonic (MS) waves (e.g., Kulsrud \& Ferrari 1971), 
and Alfv\'en waves (e.g., Miller \& Roberts 1995).

It has been shown that re--acceleration of a population of relic 
electrons by turbulence powered by major mergers is suitable to explain 
the very large scale of the observed radio emission and may also 
account for the complex spectral behaviour observed in some 
diffuse radio sources (Brunetti et al., 2001; 
Petrosian 2001; Ohno, Takizawa and Shibata 2002; 
Fujita, Takizawa and Sarazin 2003).
These first calculations, however, did not take into account the
full coupling between particles and MHD waves (Alfv\'en and MS)
and consider only relativistic electrons neglecting the presence
of the relativistic hadrons in the ICM.
In the following we focus on the results from recent calculations 
of fully coupled and 
time-dependent Alfv\'enic acceleration in the ICM.
A first application to the case of MS waves 
in the ICM including the relevant coupling and damping processes 
can be found in Cassano \& Brunetti (2004)
(see also Cassano \& Brunetti, these proceedings).

\begin {figure*}[t]
\vskip 0cm
\centerline{\epsfysize=8cm\epsfbox{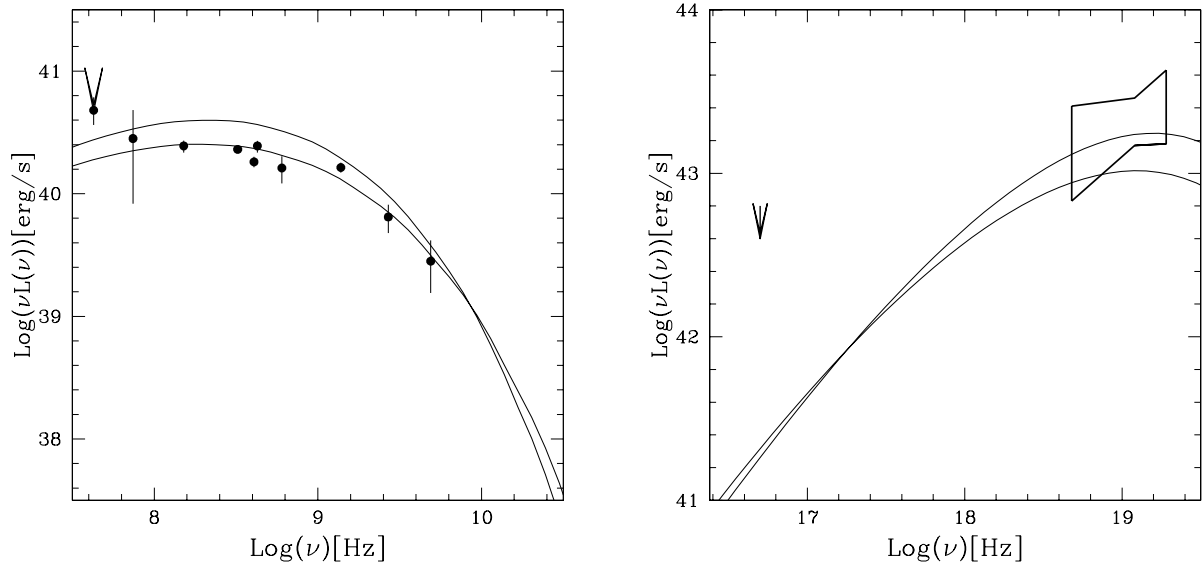}}
\vskip -0.3cm
\caption{SYN and IC spectra calculated for a Coma--like
cluster and compared with the radio, EUV (considered as
an upper limit, see Brunetti et al.~2004) and HXR data.
Calculations are obtained for a central field $B_o=1.5 \mu$G,
$B(r)\propto n_{\rm th}^{2/3}$, an injection power of Alfv\'en
waves $P_{\rm A} \propto n_{\rm th}^{5/6}$ (e.g., Brunetti
et al.~2004), $X_{\rm th}^{\rm p}$=const, 
and an initial electron and proton energy 
density $= 5\times 10^{-5}$ and $10^{-2}$ that of the 
thermal pool. For the reported models
the energy of the fluid turbulence
is in the range 15--30\% of the thermal pool. 
}
\vskip -0.2cm
\end{figure*}

\begin {figure*}[t]
\vskip 0cm
\centerline{\epsfysize=8cm\epsfbox{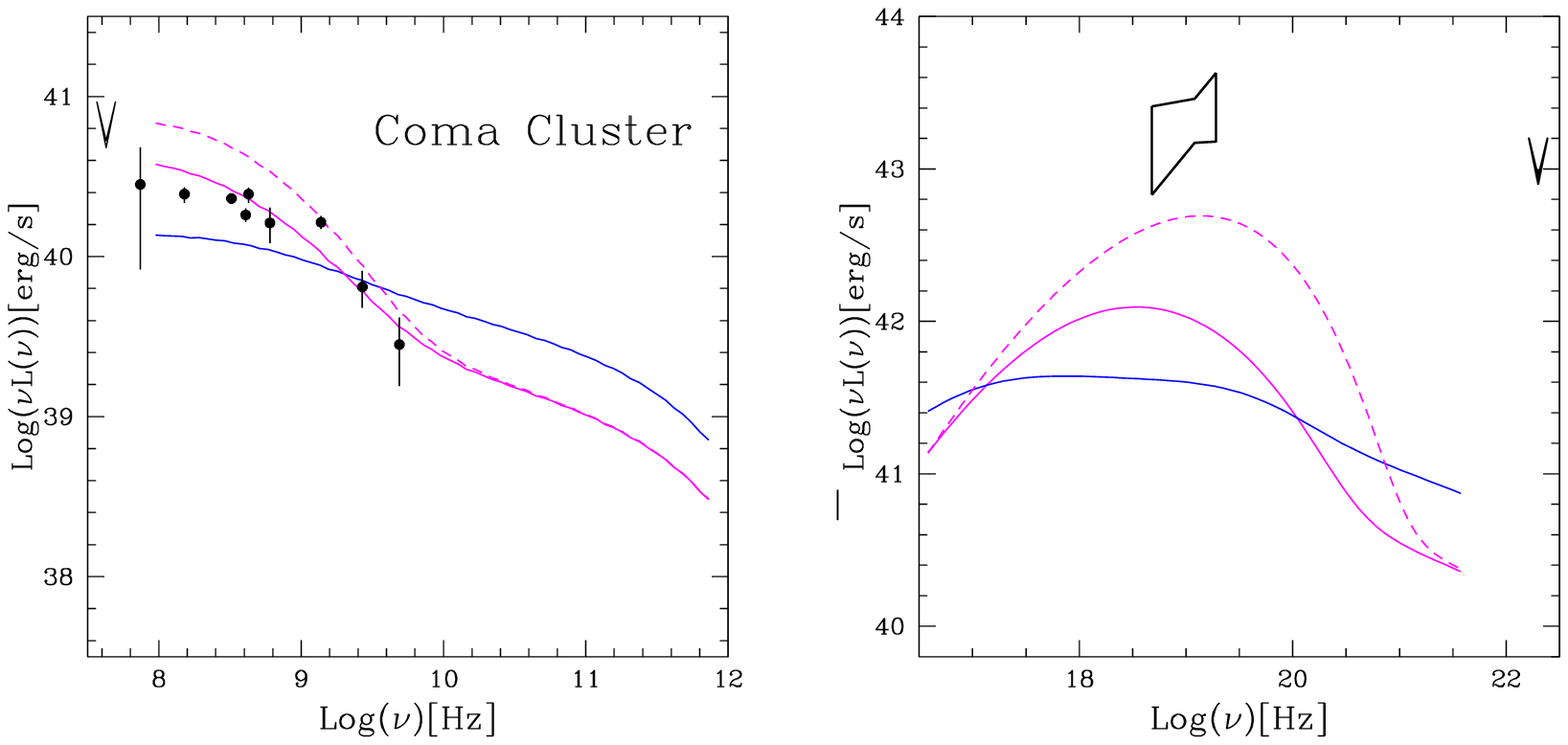}}
\vskip -0.3cm
\caption{SYN and IC spectra calculated for a Coma--like
cluster and compared with the radio, HXR and gamma--ray
data.
Calculations are obtained for a central field $B_o=1.5 \mu$G,
$B(r)\propto n_{\rm th}^{2/3}$, $X_{\rm th}^{\rm p}$=const,
and an injection power of Alfv\'en
waves $P_{\rm A} \propto n_{\rm th}^{5/6}$ (solid lines) 
and $P_{\rm A} \propto n_{\rm th}^{5/6}$ (dashed lines).
The initial proton energy is $3\times 10^{-2}$ (magenta lines)
and $10^{-1}$ (blue lines) that of the
thermal pool.
No {\it relic}--primary electrons are considered in the
calculations.
For the reported models the energy of the fluid turbulence
is in the range 20--70\% of the thermal pool (70\% for
dashed models). Calculations are stopped at $\sim 3\times 
10^{21}$Hz in the right panel.
}
\vskip -0.2cm
\end{figure*}

\subsection {Acceleration of Protons and Primary Electrons}

Very recently, the problem of particle-Alfv\'en wave interactions has
been investigated in the most general situation in which relativistic 
electrons (primary), 
thermal protons and relativistic protons exist within the 
cluster volume (Brunetti et al., 2004). 
In this modelling the interaction of all these components with the waves, 
as well as the turbulent cascading and damping processes of Alfv\'en waves, 
have been treated in a fully time-dependent way in order to 
calculate the spectra of electrons, protons and waves at any fixed time. 

The injection process of Alfv\'en waves in the ICM 
is the major hidden ingredient in these calculations
since it depends on a number of unknown quantities and 
is not a well established mechanism.
These waves couple with relativistic electrons (and protons) at very
small scales $\lambda \sim 2\pi p/(\Omega m)$ and thus 
the cascading process should be very efficient if these
waves are injected at large scales.
However, it has been shown that the cascading process of Alfv\'en waves
ends up with a highly anisotropic wave--spectrum which
strongly reduce the efficiency of particle acceleration 
(Yan \& Lazarian 2004 and ref. therein).
One possibility to obtain an efficient coupling of Alfv\'en waves
with relativistic particles is thus to inject these waves 
directly at relatively small scales in order to avoid strong
anisotropy during the cascading process.
This may be achieved via the {\it Lighthill} mechanism 
(e.g., Kato 1968; Eilek \& Henriksen 1984) which essentially 
can be adopted to convert a fraction of the energy 
flux of the cascade of the large scale fluid turbulence 
into radiation of Alfv\'en waves 
at smaller scales (Fujita et al., 2003; Brunetti et al., 2004).

Provided that this mechanism is at work in the ICM, 
Brunetti et al.(2004) have shown that an efficient particle 
acceleration is obtained in the ICM if cluster mergers
inject a fraction of large scale turbulence with energy
$\sim 10-30 \%$ of that of the thermal pool.
On the other hand, since the major damping of the Alfv\'en waves
is due to relativistic hadrons in the ICM, 
these authors have shown that efficient 
electron acceleration can be powered only by assuming that
the energy budget of relativistic protons is less than a few
percent of the thermal pool.
Under these conditions these calculations have proved that 
the observations described
in Sect.~2 (including the fine radio properties and HXR tails)
can be successfully reproduced.
In Fig.~3 we report an example of results 
obtained for of a Coma--like cluster.

\subsection {Hybrid Models}

In the most general situation, relativistic hadrons in the
ICM may store 
an appreciable fraction
of the thermal energy, and they should inject a non negligible
component of secondary electrons in the cluster volume.
The bulk of this component is associated to $\gamma \sim 100-300$ 
electrons. These electrons 
cannot be responsible for the observed synchrotron 
radio spectra, but they may be efficiently re--accelerated 
at higher energies giving a novel population of radio emitting
particles.

Very recently, we have developed Hybrid Model calculations, 
which include, in a self--consistent way,
this component togheter with the MHD waves and 
the primary electrons and protons.
We find that the 
re--acceleration of secondary electrons is a very efficient 
mechanism since these electrons are re--accelerated 
and continuously injected at the same time; 
the efficiency increases toward the central regions of the 
clusters where
the number density of the target--thermal protons is
larger.

Fig.~4 shows an example of the
results obtained from our calculations in the
case of a Coma--like cluster.
The following items should be outlined :

\begin{itemize}

\item[{\it i)}] with increasing the energy budget of relativistic
protons the efficiency of electron acceleration decreases.
This effect is not compensated by the increase of
the number of injected secondary electrons and consequently
the synchrotron spectrum at $\sim$GHz frequencies drops down.
On the other hand, the high frequency synchrotron 
tail expected at high frequencies ($\geq 10$ GHz), which is 
essentially produced by the injected fresh 
electrons, scales with the energy of the relativistic protons;

\item[{\it ii)}] an energy budget of relativistic protons
$< 10\%$ of the thermal pool is required to reproduce
a clear steepening of the synchrotron spectrum at higher
frequencies;

\item[{\it iii)}] the IC spectrum produced without introducing
a population of 
{\it relic}--primary electrons is below the HXR tail
detected by BeppoSAX.
This is simply because the number of secondary electrons
in the cluster volume is limited by the decrease of the
number density of thermal protons in the external regions.
At least in the external regions, a 
population of {\it relic}--primary electrons (to be reaccelerated)
is thus required to match the HXR;

\item[{\it iv)}] similarly to the case of secondary models,
gamma ray emission is expected by 
Hybrid Models.
In Fig.~4 we report only the contribution from IC, while
an additional contribution is expected by the decay of the
$\pi^o$ produced during the hadronic collisions.
However, at variance with secondary models, the amount
of synchrotron GHz emission (and HXRs) 
is basically 
anti--correlated with the strength of the gamma ray
emission.

\end{itemize}

A more detailed discussion will be presented
in a forthcoming paper (Brunetti \& Blasi, in prep.).

\section {CONCLUSIONS}
The fine radio properties of radio halos and their statistics 
are a powerful tool to understand the origin of the emitting
electrons.
The study of the spectrum of radio halos and their
very broad extension are crucial tests for the proposed models.
In particular, the spectral steepenings observed in a few radio
halos indicate the presence of a high energy cut--off
in the spectrum of the emitting electrons at about $\gamma \sim
10^4$; we have shown that these steepenings are not
appreciably affected by the SZ correction.
The broad synchrotron profiles of some giant and luminous
radio halos (e.g., Abell 2163) are also a challenge for secondary
models: in order to mantain the energy budget of the relativistic
hadrons significantly below that of the thermal pool, 
the strength of the magnetic field in the ICM should
remain almost constant on scales comparable to the cluster size,
which contrasts with the scenario of the amplification of
the magnetic fields in clusters.
A similar energetic problem is also found by looking at the
spectral indices of a sample of luminous radio halos
(Sect.~2b, Fig.~1).

All the above considerations suggest that the process of
injection of secondary electrons in the ICM may not be
the leading mechanism in producing the observed non--thermal
activity.
Thus we have focussed on {\it in situ} particle acceleration 
models and discussed a class of self--consistent and time--dependent
models which follow the interaction between MHD waves, protons
(relativistic and thermal) and electrons (relativistic).
The major hidden ingredient in these models is the injection
of turbulence in the ICM and the generation of Alfv\'en and 
Fast MS waves.
In this paper we have focussed on Alfv\'en waves which damps most
of their energy flux on the relativistic protons in the ICM.
In this case the electron acceleration is thus limited
by the presence of
the relativistic hadrons: efficient electron acceleration is
obtained only provided that hadrons store a few percent
of the thermal energy.
Under these conditions, 
we have shown that if a fraction of $\sim 10-30 \%$ of the energy
of the thermal cluster is in the form of turbulence, 
then particle acceleration is found to be sufficiently
efficient to generate giant radio halos (and possibly HXR tails).
A nover approach is provided by Hybrid Models in which 
both primary and secondary electrons are considered.
Although the reacceleration of only secondary electrons may reproduce
the observed radio properties, it is found that a population
of {\it relic}--primary electrons in the external regions (at least)
is necessary to explain the HXR tails.

\acknowledgements{I thank L.Feretti for providing the data in
Figs.~1 \& 2. This research is partially supported by
INAF (Istituto Nazionale di Astrofisica) 
through grant D4/03/15.}

\end{document}